\documentclass{article}

\usepackage{PRIMEarxiv}
\usepackage[utf8]{inputenc} 
\usepackage[T1]{fontenc}    
\usepackage{hyperref}       
\usepackage{url}            
\usepackage{booktabs}       
\usepackage{amsfonts}       
\usepackage{nicefrac}       
\usepackage{microtype}      
\usepackage{lipsum}
\usepackage{graphicx}
\usepackage{amsmath}
\usepackage{comment}
\graphicspath{{media/}}     

\title{A Model-Based Synthetic Stock Price Time Series Generation Framework}

\author{
    Haibei Zhu, Svitlana Vyetrenko, Tucker Balch\\
    J.P. Morgan AI Research\\
    383 Madison Avenue, New York, NY 10017, USA\\
    \texttt{\{haibei.zhu, svitlana.s.vyetrenko, tucker.balch\}email@jpmchase.com} \\
}

\begin{document}
\maketitle

\begin{abstract}
The Ornstein-Uhlenbeck (OU) process, a mean-reverting stochastic process, has been widely applied as a time series model in various domains. This paper describes the design and implementation of a model-based synthetic time series model based on a multivariate OU process and the Arbitrage Pricing Theory (APT) for generating synthetic pricing data for a complex market of interacting stocks. The objective is to create a group of synthetic stock price time series that reflects the correlation between individual stocks and clusters of stocks in how a real market behaves. We demonstrate the method using the Standard and Poor's (S\&P) 500 universe of stocks as an example.
\end{abstract}


\section{Introduction}
\label{ch_introduction}

The dynamic nature of financial markets, represented by the fluctuation of stock prices, is in response to various factors ranging from political events to company news \cite{cutler1988moves}. While the price time series seem random, they follow specific underlying patterns and dependencies that can be captured and modeled \cite{lebaron1999time}. The Ornstein-Uhlenbeck (OU) process, a mean-reverting stochastic process focusing on modeling the trend and variance of time series, is an appropriate model-based technique to model price time series \cite{uhlenbeck1930theory, byrd2019explaining}. Given the complexity of financial markets, traditional univariate models might often fall short in encapsulating the nuances present. In light of this, this paper extended the univariate OU process into a multivariate model to capture the properties of individual time series and the interrelation among time series. By incorporating multiple dimensions into the model, we aim to ensure that the dependencies and correlations between financial price time series are captured. In this paper, we first generated synthetic market sector ETFs, which track representative stocks specific to an industry sector. We passed the ETF time series to the Arbitrage Pricing Theory (APT) framework to generate synthetic individual stocks in the Standard and Poor's (S\&P) 500.

As the core of the OU process, the mean-reverting stochastic process diverges from a simple random walk by possessing the tendency to revert to a long-term mean \cite{poterba1988mean}. Utilizing such a property, we assume that stock price time series tend to return to a long-term average or trend despite short-term fluctuations. However, financial markets are seldom about isolated events or individual stocks, and they represent complex interactions and correlations \cite{pollet2010average, preis2012quantifying}. The S\&P 500, which includes the stocks from 500 large-cap companies, is a typical example of a time series dataset with such interrelations. Modeling the complexity requires understanding the dependencies and correlations between the time series. This is where the nature of the multivariate OU process is specialized.

Synthetic data, especially when rooted in robust modeling, has significant applications like modeling extreme or unseen market scenarios, training machine learning models in a controlled environment, and extending existing datasets \cite{assefa2020generating}. The primary objective of this study is to develop a model-based synthetic stock price time series data generation framework and to demonstrate a synthetic time series example generated from this framework. This paper is structured as follows. Section \ref{ch_multi_ou} covers the detailed equation of the multivariate OU process and demonstrates the generation process of the market sector ETF time series. This section also presents the preliminary evaluation of the generated time series. Section \ref{ch_apt} introduces the APT framework and presents the synthetic stock data. We discuss the approach and results in Section \ref{ch_conclusion} and conclude our work.

\begin{figure}
    \centering
    \includegraphics[width=\columnwidth]{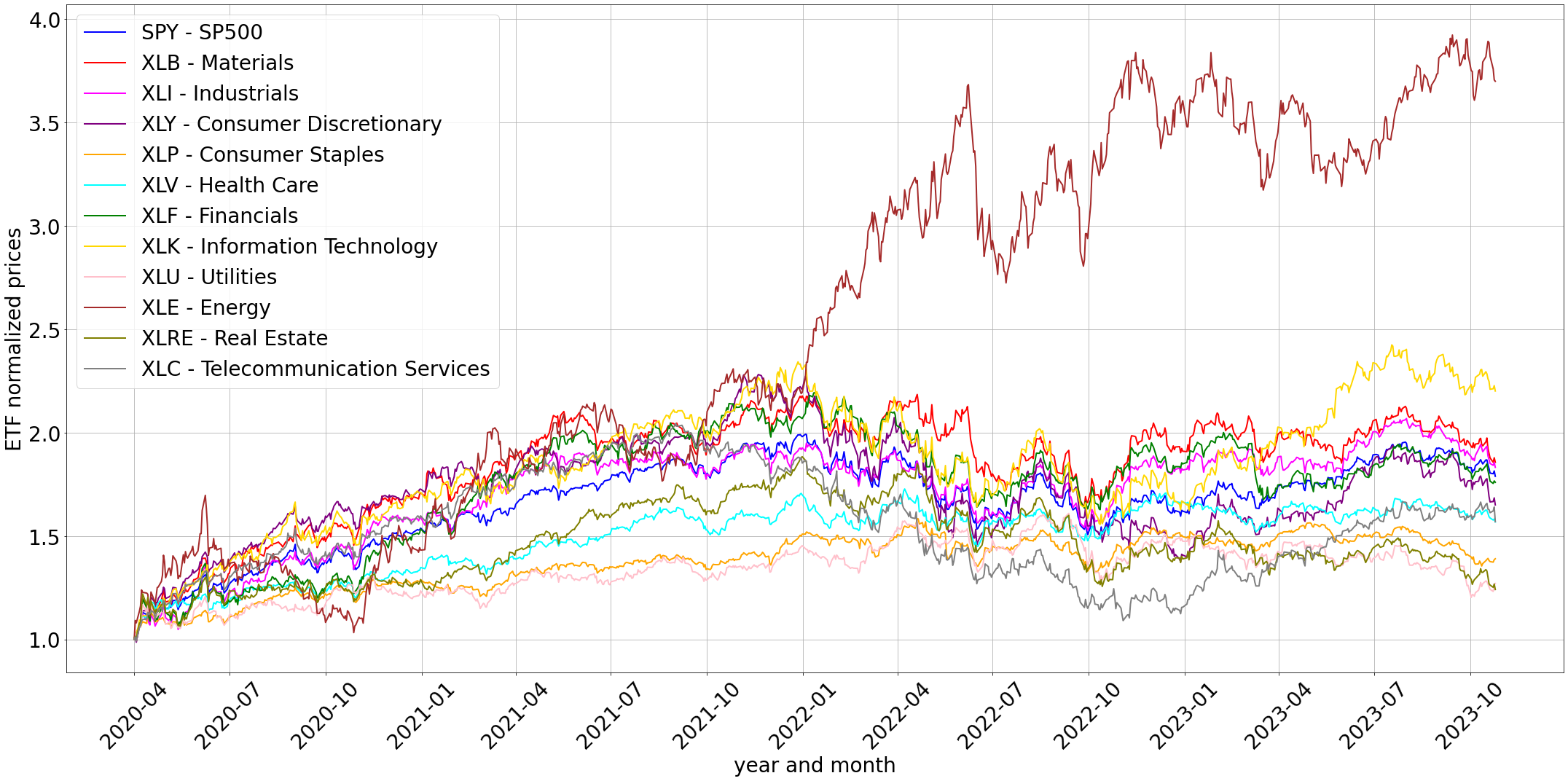}
    \caption{Normalized historical sector ETF price time series.}
    \label{fig_real_etf}
\end{figure}

\section{Multivariate Ornstein-Uhlenbeck (OU) Process}
\label{ch_multi_ou}

\subsection{Formulation}

The intricacies of financial markets demand models capable of capturing multiple time series' simultaneous evolution and interdependencies. The multivariate OU process precisely adapts to this requirement with its capability of handling both the deterministic (mean-reversion) and the stochastic aspects of price time series. Equation \eqref{eq_multi_ou} encapsulates the essence of the multivariate OU process, as shown below:
\begin{equation}
\begin{aligned}
x_{i,t}-x_{i,t-1}=\Delta x_{i,t}&=A\cdot(\mu_i-x_{i,t-1}+\gamma_i\cdot t)+\Sigma \\
\begin{bmatrix}
\Delta x_{1,t} \\[4pt]
\Delta x_{2,t} \\[4pt]
... \\[4pt]
\Delta x_{N,t} \\
\end{bmatrix}&=
\begin{bmatrix}
\theta_{1,1},\theta_{1,2},...,\theta_{1,N} \\[4pt]
\theta_{2,1},\theta_{2,2},...,\theta_{2,N} \\[4pt]
... \\[4pt]
\theta_{N,1},\theta_{N,2},...,\theta_{N,N} \\
\end{bmatrix}\cdot
\begin{bmatrix}
\mu_1-x_{1,t}+\gamma_1\cdot t \\[4pt]
\mu_2-x_{2,t}+\gamma_2\cdot t \\[4pt]
... \\[4pt]
\mu_N-x_{N,t}+\gamma_N\cdot t \\
\end{bmatrix}+
\begin{bmatrix}
\sigma_{1,1},\sigma_{1,2},...,\sigma_{1,N} \\[4pt]
\sigma_{2,1},\sigma_{2,2},...,\sigma_{2,N} \\[4pt]
... \\[4pt]
\sigma_{N,1},\sigma_{N,2},...,\sigma_{N,N} \\
\end{bmatrix}
\end{aligned}
\label{eq_multi_ou}
\end{equation}
The central premise of the multivariate OU process, as depicted by the equation, is the modeling of value changes ($\Delta x_{i,t}$, which is the difference between $x_{i,t}$ and $x_{i,t-1}$) in a given time series based on the adjacent values and the deviation from its long-term mean or trend. Specifically, in Equation \eqref{eq_multi_ou}:
\begin{itemize}
    \item \textbf{Reversion to the mean}: The term $(\mu_i-x_{i,t-1}+\gamma_i\cdot t)$ represents the deviation of the $i^{th}$ time series from its long-term mean $\mu_i$, adjusted by a step-wise linear trend $\gamma_i$. This deviation is the driving force pushing the time series back towards its mean. Here, $x_{i,t}$ is the $t^{th}$ data point on the $i^{th}$ dimension. The total dimension, or the number of time series, is $N$. $\mu_i$ is the long-term mean of the time series, and $\gamma_i$ represents the step-wise trend constant. This formulation captures the essence of mean-reversion, the characteristic of the OU process.
    \item \textbf{The matrix of reversion rates}: This matrix, $A$, with individual elements $\theta_{i,j}\;(i,j\in [1,N])$, modulates the speed in which each time series reverts to its mean. A larger value of $\theta_{i,j}$ implies a faster reversion for the $i^{th}$ dimension based on the $j^{th}$ dimension. The diagonal $\theta s$ represents the intrinsic reversion rate of each dimension to its mean, while the off-diagonal $\theta s$ captures the influence of one dimension on the reversion rate of another. In a multidimensional time series system, these interactions can shed light on the underlying dependencies and couplings between different time series dimensions.
    \item \textbf{Stochastic term}: $\Sigma$ Representing the inherent randomness of financial markets, this term introduces volatility and uncertainty into the model. The covariance matrix, $\Sigma$, captures the co-movements and shared volatility structure among the various time series.
\end{itemize}

\begin{figure}
    \centering
    \includegraphics[width=\columnwidth, height=0.44\textheight]{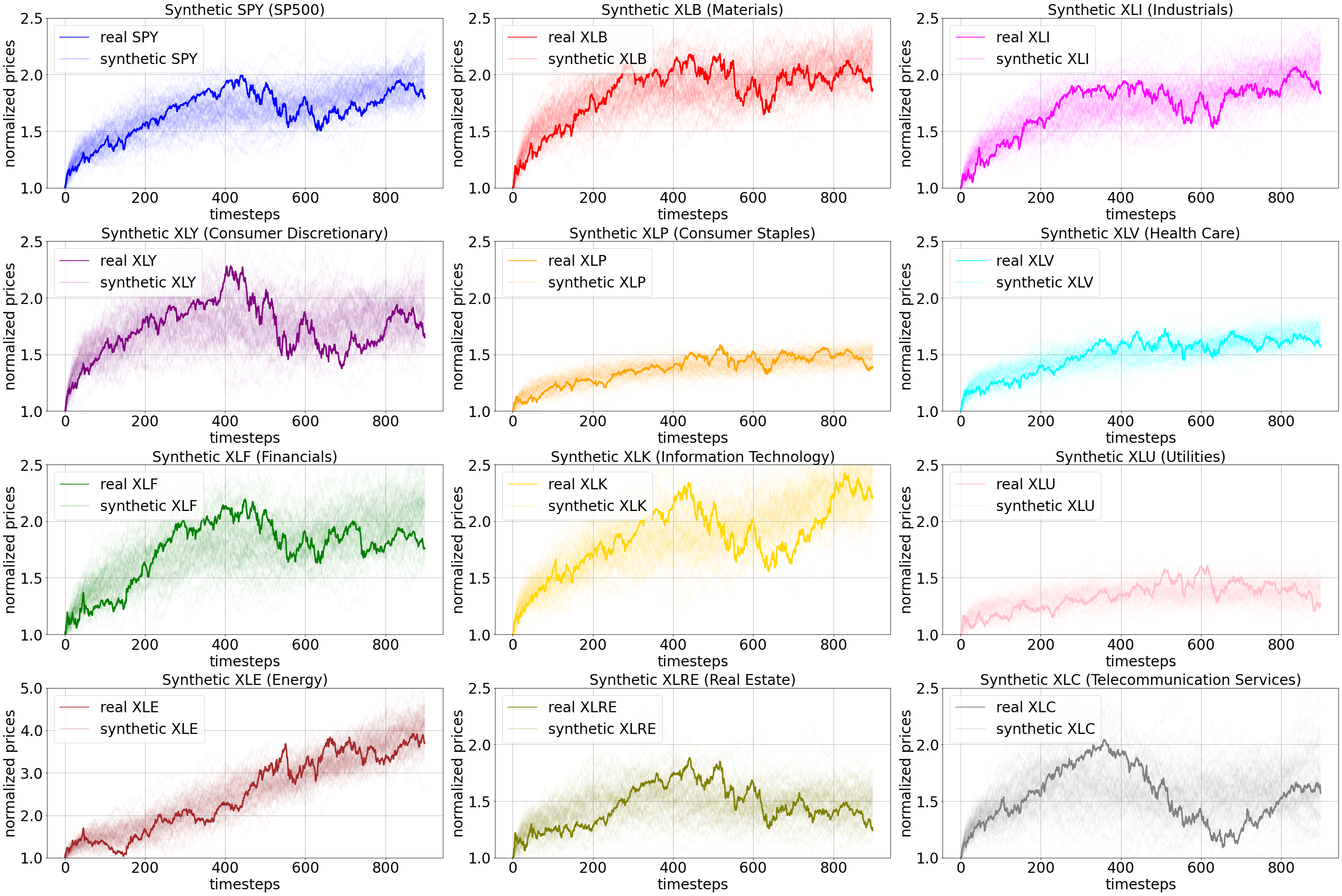}
    \caption{Multiple generations of normalized synthetic ETF prices time series.}
    \label{fig_syn_etf_multi_trace}
\end{figure}

\subsection{Synthetic Market Sector ETF Time Series Generation}

The S\&P 500 provides a broad snapshot of the economic landscape by encapsulating the performance of 500 leading large-cap companies and spans multiple sectors. Various Exchange Traded Funds (ETFs) have been developed to provide more insights into distinct economic segments. We consider the ETFs as the factors for the APT framework, elaborated in the following section, because of the ETFs' diversification and breadth presentation. Sector ETFs represent a diversified basket of stocks within a particular sector, reducing unsystematic risk and allowing a focus on systematic factors. Thus, we can isolate and analyze the sensitivities of each sector to individual assets. Also, by examining sector ETFs, we can observe a broader representation of the economy and the different dynamics.

Specifically, the SPY ETF offers a comprehensive view, tracking the overall performance of the entire S\&P 500 index. Figure \ref{fig_real_etf} presents the normalized prices of the SPY and 11 sector ETFs. The starting price for all ETFs is set to 1, and the starting date is $04/01/2020$, which is the start of the second quarter of 2020. The normalized prices are propagated based on the starting price and the real price returns. This normalization allows for easier comparison of relative changes across different ETFs over time. We utilized the multivariate OU process in this data generation framework to generate synthetic sector ETFs time series.

\subsubsection{Multivariate OU Parameter Estimation}

Other than the normalized price for SPY, the S\&P index, for the other 11 sector ETFs, we used the relative normalized prices as the time series for estimating multivariate OU parameters. The relative normalized prices are propagated based on relative returns, which are the return of a sector ETF subtracted from the corresponding return of the market, the SPY return. In this case, we can isolate the impact from the market and only concentrate on the changes for each sector. A linear regression model is employed to fit the ETF time series:
\begin{equation}
\Delta x=\theta\cdot x+\epsilon
\label{eq_linear}
\end{equation}
Here, $\Delta x$ represents the change between adjacent time series values, and $x$ represents the previous time series value and a timestamp. The coefficient $\theta$ represents the rate of reversion, which is the speed at which the series reverts to its mean. From this linear regression model, multiple outputs are estimated for the multivariate time series, including a rate of reversion matrix, a list of trend factors, a list of mean values that indicate the long-term expectation of the prices, and a covariance matrix calculated from the residuals from the regression fitting.

\begin{figure}
    \centering
    \includegraphics[width=\columnwidth]{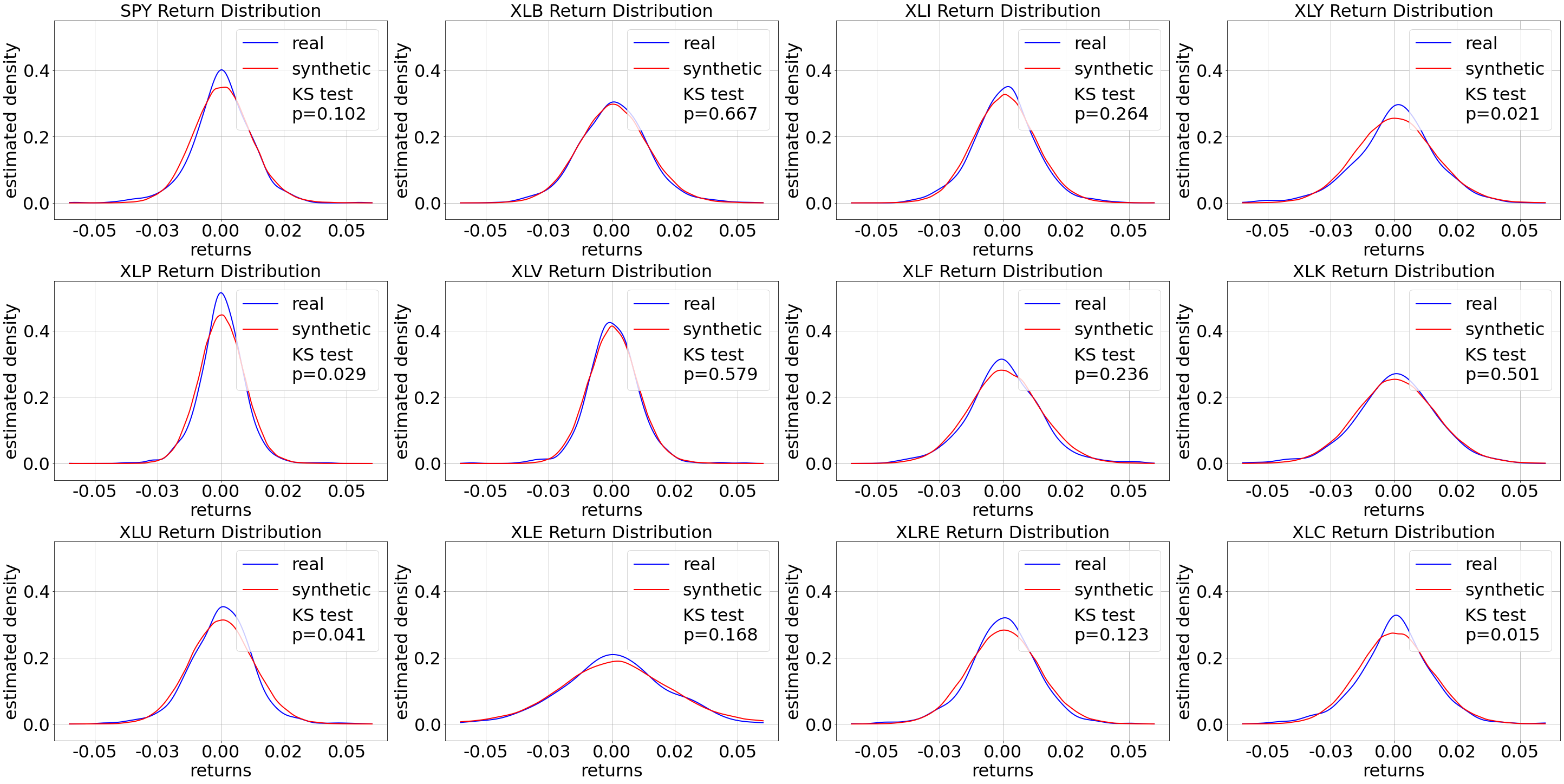}
    \caption{Return distribution comparisons between the real and synthetic time series.}
    \label{fig_return_dist}
\end{figure}

\subsubsection{Multivariate OU Time Series Generation}

The parameters obtained from the estimation process are then used to simulate the multivariate OU process and generate multivariate time series data using Equation \eqref{eq_multi_ou}. The multivariate OU process will simultaneously generate time series data in all dimensions, including the normalized price of SPY and the relative normalized prices for the other ETFs. The transformation from relative normalized prices to normalized prices is straightforward -- we first calculated the relative returns from the relative normalized prices, added the corresponding market/SPY returns to obtain sector ETF returns, and propagated the normalized prices.

An example of multiple-trace generation for all the ETFs is shown in Figure \ref{fig_syn_etf_multi_trace}. The real ETF normalized price time series is highlighted in each subplot. These synthetic time series mirror the real data regarding general trends and volatility. Each trace for a specific ETF represents a possible scenario reflecting the market dynamics. The variations of the traces are not just noise but structured deviations that embody the stochastic nature of financial markets. This feature is particularly relevant when researchers aim to generate market scenarios.

\subsubsection{Synthetic Time Series Evaluation}

\begin{figure}
    \centering
    \includegraphics[width=\columnwidth]{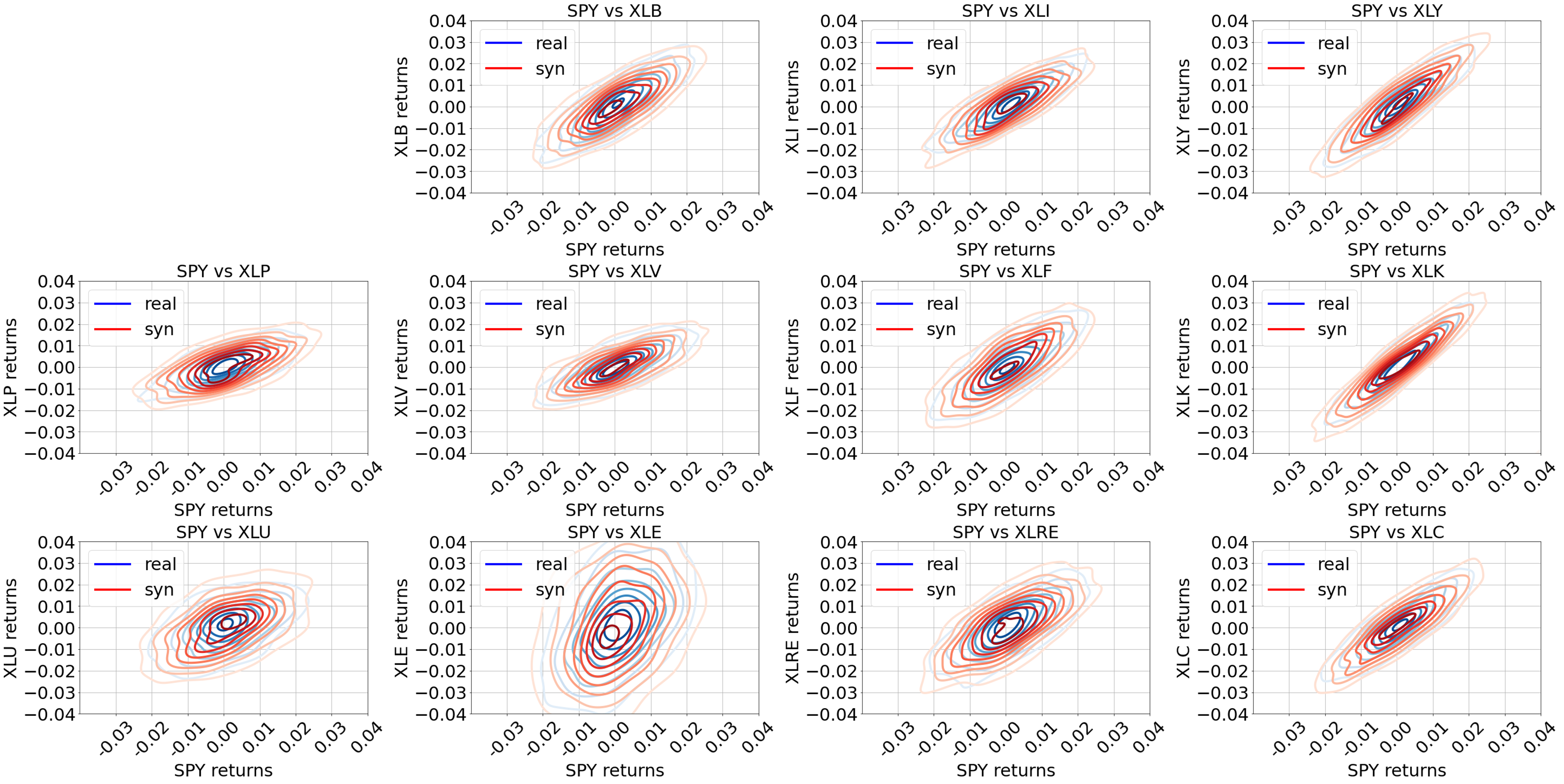}
    \caption{Joint return distribution comparisons between the real and synthetic time series.}
    \label{fig_joint_return}
\end{figure}

The effectiveness of the synthetic time series model lies not just in its capacity to mirror real data but also in its ability to encapsulate the original data's dynamics, patterns, and characteristics. In this subsection, we evaluate the generated synthetic ETF time series against the actual time series to understand the usability of the synthetic data. The initial evaluation involved plotting the synthetic time series alongside the actual time series for visual assessment. Figures \ref{fig_real_etf} and \ref{fig_syn_etf_multi_trace} provided qualitative visualization insights into how closely the synthetic data mirrored the real trends. Descriptive statistics were computed for real and synthetic datasets, shown in Figure \ref{fig_return_dist} and \ref{fig_joint_return}.

We first plotted the Kernel Density Estimation (KDE) curves for the return distributions from the real and synthetic data for the 12 ETFs in Figure \ref{fig_return_dist}. The real returns are illustrated with blue curves, while the synthetic returns are depicted with red curves. Each subplot represents the estimated density of returns for a particular ETF. These KDE plots offer insights into the distribution's shape, spread, and central tendency. A closer overlap between the blue and red curves suggests that the synthetic data closely replicates the distribution of the real data. To quantitatively measure the overlap, each plot displays the result of a Kolmogorov-Smirnov (KS) test, a non-parametric test used to compare two distributions. The p-values of the KS tests are shown in each plot to indicate the level of similarity in the distributions between real and synthetic data.

For example, the KS test for SPY return distributions has a p-value of 0.102, which means that there is not enough evidence to claim that the real and synthetic distributions are different. In contrast, the KS test for XLP has a p-value of 0.029, suggesting a significant difference between the distributions. Among these 12 entities, 8 ETFs show a closer match between real and synthetic distributions with p-values greater than the significance level of 0.05. Others, where p-values are less than the significance level and indicate challenges in replicating the real distribution, still show close alignment visually between the KDE curves. The results from the comparisons of return distributions point towards potential refinements and improvements in future iterations of the model to enhance its fidelity.

We also plotted the joint distribution KDE contours between real and synthetic data, as shown in Figure \ref{fig_joint_return}. We plotted and analyzed joint distributions between SPY and the other 11 sector ETFs because the correlation between sector ETFs and the whole market, SPY, has been widely studied and can provide insights into how the ETFs interact. In Figure \ref{fig_joint_return}, each subplot represents a bivariate distribution of two entities. The contours for the real dataset are represented in blue, while those for the synthetic dataset are in red.

In areas where the blue and red contours overlap, it indicates that the real and synthetic data distributions are similar in those regions. The higher the degree of overlap, the closer the synthetic data replicates the distribution of the real data. Some subplots, like SPY versus XLB and XLK, show a high degree of overlap, suggesting that the synthetic data generation process has effectively replicated the joint distribution of the real data. Some other subplots, like SPY versus XLU and XLE, where there is less overlap in the contour shapes, indicate areas where the synthetic data might not closely match the real data regarding joint distribution. Such results also provide a reference, or benchmark, for refining and improving future data generation techniques.

\begin{figure}
    \centering
    \includegraphics[width=\columnwidth]{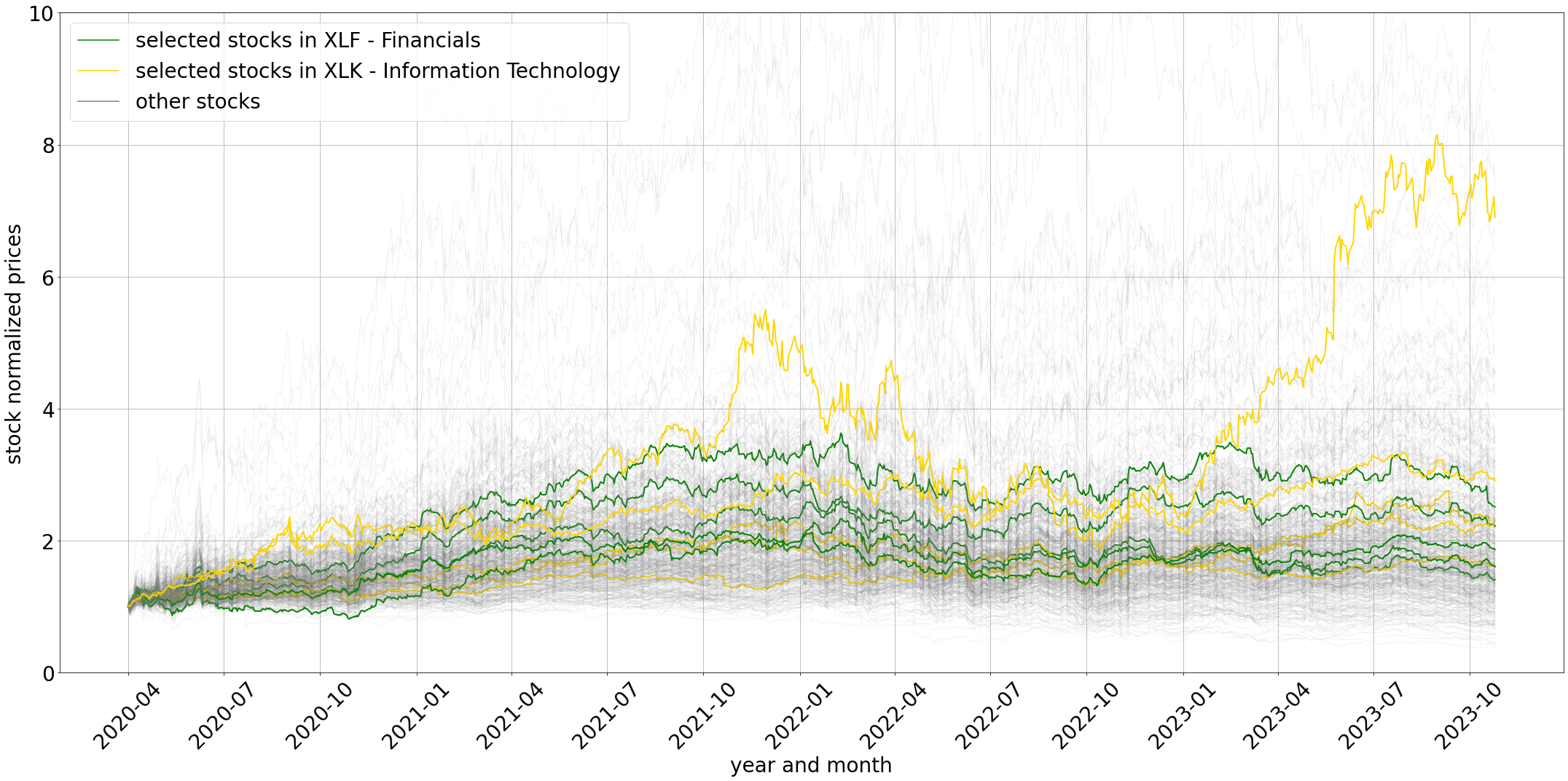}
    \caption{Normalized historical stock prices time series. Several selected stocks in XLF and XLK are highlighted.}
    \label{fig_real_stock}
\end{figure}

\section{Arbitrage Pricing Theory (APT)}
\label{ch_apt}

The APT framework is a multi-factor asset pricing model that captures the relationship between a financial asset's returns and the macroeconomic factors that potentially influence it \cite{ross2013arbitrage, roll1980empirical}. Unlike the Capital Asset Pricing Model (CAPM) \cite{blume1973new, fama2004capital}, which suggests a single-factor model based on market risk, APT assumes that multiple factors could affect returns. The expected return of an asset is modeled as a linear combination of various macroeconomic factors or theoretical market indices, with sensitivity coefficients for all factors and a stock-specific risk. In this work, we utilized the return of the market index, SPY, and the relative returns from the 11 sector ETFs as the systematic factors in APT.

\subsection{Formulation}

The following equation can represent the mathematical foundation of the APT:
\begin{equation}
r_i(t)=\alpha_i+\beta_{i1}f_1(t)+\beta_{i2}f_2(t)+\dots+\beta_{in}f_n(t)+\epsilon_i(t)
\label{eq_apt}
\end{equation}
In this equation:
\begin{itemize}
    \item $r_i(t)$ is the return on the $i^{th}$ asset on day $t$.
    \item $\alpha_i$ is the expected return on asset $i$ that is unrelated to the factor exposures.
    \item $\beta_{in}$ is the sensitivity, or the weight coefficient, of the $i^{th}$ asset to factor $n$.
    \item $f_n(t)$ is the value of the $n^{th}$ systematic factor on day $t$.
    \item $\epsilon_i(t)$ is the idiosyncratic risk of asset $i$, or the component of the return on asset $i$ that is not explained by the factor exposures.
\end{itemize}

\begin{figure}
    \centering
    \includegraphics[width=\columnwidth]{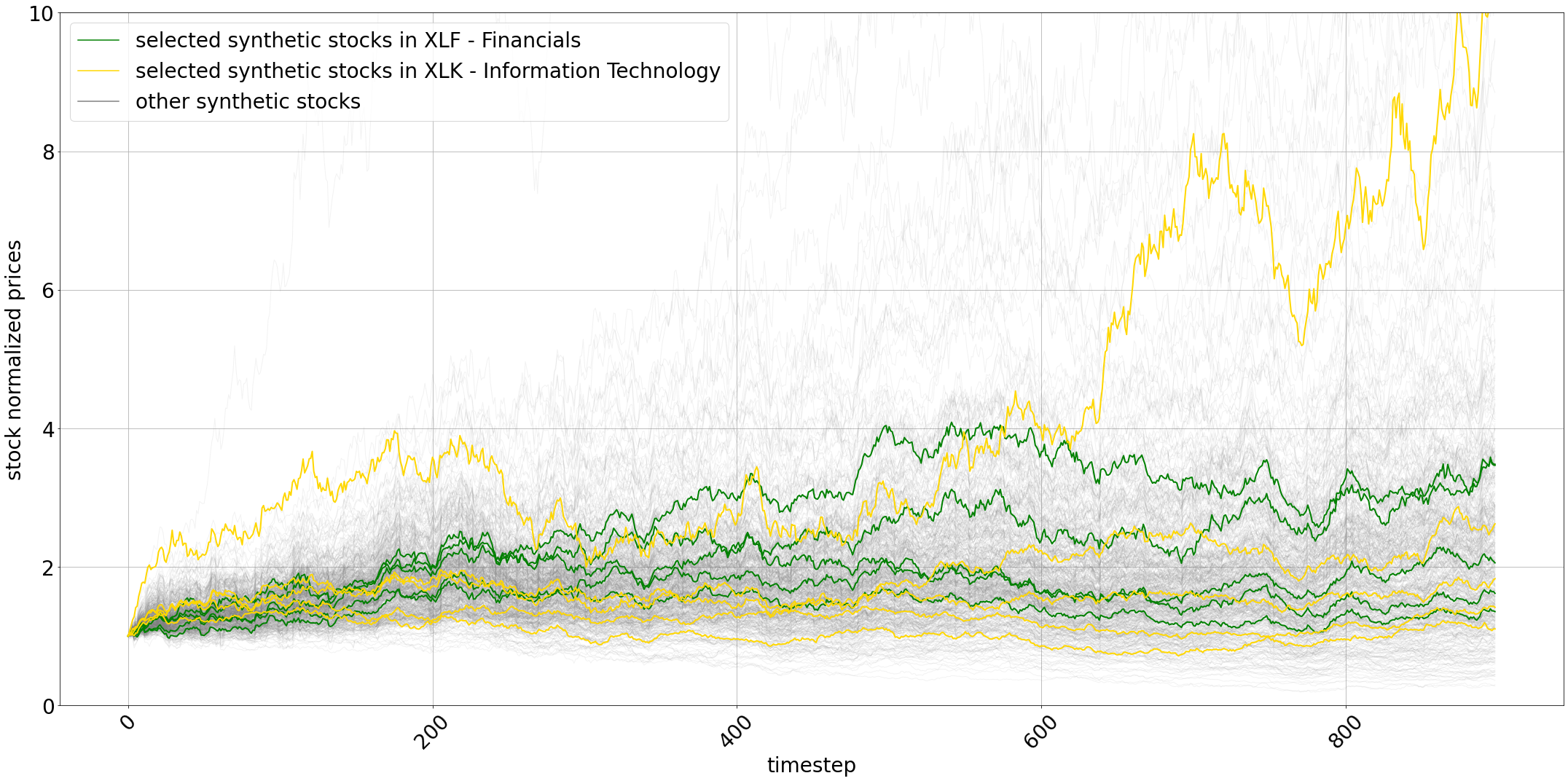}
    \caption{Synthetic normalized stock prices time series.}
    \label{fig_syn_stock}
\end{figure}

\subsection{Synthetic Stock Price Time Series Generation}

We first estimated the coefficients, $\beta_{in}$, the constant, or the expected return, $\alpha_i$, and the residuals in the linear model, $\epsilon_i(t)$, for all stocks using Equation \eqref{eq_linear} given time series from the historical ETFs and stocks. This step is similar to the parameter estimation for the multivariate OU process, as mentioned in Section \ref{ch_multi_ou}. Then, we utilized the APT framework to generate individual stock price time series. Specifically in Equation \eqref{eq_apt}, $f_1(t)$ is the return of the market index, SPY, on the corresponding date, and $f_n(t), n\neq 1$ are the relative returns from the other 11 sector ETFs. All $f_n(t)$ values are calculated from synthetic ETF time series, whose generation process is also mentioned in Section \ref{ch_multi_ou}.

A historical stock price time series example is shown in Figure \ref{fig_real_stock}, in which several selected stocks in the XLF sector are highlighted in green, several stocks in the XLK sector are highlighted in yellow, and the rest of the stocks in S\&P are shown in light grey. As a comparison, a synthetic version of the stocks is shown in Figure \ref{fig_syn_stock}. In the historical time series, both the selected stocks in XLF and XLK show a pronounced upward trend starting from around early 2021, with XLK experiencing more growth and reaching its peak around early 2023. The synthetic stock time series portrays a similar trajectory, with the selected synthetic stocks in both XLF and XLK sectors showing a noticeable rise, especially the XLK. It is noteworthy that the stocks in XLK outperform the stocks in XLF consistently in both historical and synthetic data in terms of long-term returns, suggesting that the synthetic generation process captures sector-specific trends effectively. Also, historical and synthetic data reveal a fair amount of volatility, with multiple peaks throughout the whole time range.

The example of synthetic stocks shown in Figure \ref{fig_syn_stock} is one possible outcome of this data generation framework. Because of the randomness in the multivariate OU process, we can generate infinite sets of synthetic stocks given the historical market data. Synthetic stock data offers a valuable tool for simulating various market conditions, enabling traders and researchers to explore various hypothetical market conditions without risking actual capital.

\section{Conclusion}
\label{ch_conclusion}

This work has introduced a composite synthetic stock price generation framework that utilizes the multivariate OU process with the principles of APT. This framework can simulate the stochastic and mean-reverting behavior of stock prices while capturing systemic factors that influence market dynamics for individual stocks. This approach has the capacity to preserve the essential statistical and correlation properties inherent to real stock market data, thereby producing a high-fidelity reflection of market fluctuations.

The evaluations of our synthetic data example against historical data indicate a high level of fidelity, demonstrating the generation framework's efficacy in replicating market movements. While deviations in certain synthetic ETFs show the challenges of capturing the full spectrum of market volatility, they highlight areas for future refinement. This framework represents a baseline or benchmark in synthetic data generation for financial markets, offering a model-based tool to capture stock market dynamics and serving as a reference for further research and development in this domain.

\section*{Acknowledgments}
This paper was partly prepared for informational purposes by the Artificial Intelligence Research Group of JPMorgan Chase \& Co and its affiliates ("J.P. Morgan") and is not a product of the Research Department of J.P. Morgan. J.P. Morgan makes no representation and warranty whatsoever and disclaims all liability for the information's completeness, accuracy, or reliability. This document is not intended as investment research or advice, or a recommendation, offer, or solicitation for the purchase or sale of any security, financial instrument, financial product, or service, or to be used in any way for evaluating the merits of participating in any transaction, and shall not constitute a solicitation under any jurisdiction or to any person if such solicitation under such jurisdiction or to such person would be unlawful.

\bibliographystyle{unsrt}
\bibliography{references}

\begin{thebibliography}{10}

\bibitem{cutler1988moves}
David~M Cutler, James~M Poterba, and Lawrence~H Summers.
\newblock What moves stock prices?, 1988.

\bibitem{lebaron1999time}
Blake LeBaron, W~Brian Arthur, and Richard Palmer.
\newblock Time series properties of an artificial stock market.
\newblock {\em Journal of Economic Dynamics and Control}, 23(9-10):1487--1516, 1999.

\bibitem{uhlenbeck1930theory}
George~E Uhlenbeck and Leonard~S Ornstein.
\newblock On the theory of the brownian motion.
\newblock {\em Physical Review}, 36(5):823, 1930.

\bibitem{byrd2019explaining}
David Byrd.
\newblock Explaining agent-based financial market simulation.
\newblock {\em arXiv preprint arXiv:1909.11650}, 2019.

\bibitem{poterba1988mean}
James~M Poterba and Lawrence~H Summers.
\newblock Mean reversion in stock prices: Evidence and implications.
\newblock {\em Journal of Financial Economics}, 22(1):27--59, 1988.

\bibitem{pollet2010average}
Joshua~M Pollet and Mungo Wilson.
\newblock Average correlation and stock market returns.
\newblock {\em Journal of Financial Economics}, 96(3):364--380, 2010.

\bibitem{preis2012quantifying}
Tobias Preis, Dror~Y Kenett, H~Eugene Stanley, Dirk Helbing, and Eshel Ben-Jacob.
\newblock Quantifying the behavior of stock correlations under market stress.
\newblock {\em Scientific Reports}, 2(1):752, 2012.

\bibitem{assefa2020generating}
Samuel~A Assefa, Danial Dervovic, Mahmoud Mahfouz, Robert~E Tillman, Tucker Balch, Prashant Reddy, and Manuela Veloso.
\newblock Generating synthetic data in finance: Opportunities, challenges and pitfalls.
\newblock In {\em Proceedings of the First ACM International Conference on AI in Finance}, pages 1--8, 2020.

\bibitem{ross2013arbitrage}
Stephen~A Ross.
\newblock The arbitrage theory of capital asset pricing.
\newblock In {\em Handbook of the Fundamentals of Financial Decision Making: Part I}, pages 11--30. World Scientific, 2013.

\bibitem{roll1980empirical}
Richard Roll and Stephen~A Ross.
\newblock An empirical investigation of the arbitrage pricing theory.
\newblock {\em The Journal of Finance}, 35(5):1073--1103, 1980.

\bibitem{blume1973new}
Marshall~E Blume and Irwin Friend.
\newblock A new look at the capital asset pricing model.
\newblock {\em The Journal of Finance}, 28(1):19--33, 1973.

\bibitem{fama2004capital}
Eugene~F Fama and Kenneth~R French.
\newblock The capital asset pricing model: Theory and evidence.
\newblock {\em Journal of Economic Perspectives}, 18(3):25--46, 2004.

\end{thebibliography}

\end{document}